\documentclass[conference]{IEEEtran}
\usepackage{cite}
\usepackage{amsmath,amssymb,amsfonts}
\usepackage{algorithmic}
\usepackage{graphicx}
\usepackage{textcomp}
\usepackage{xcolor}
\usepackage{booktabs}
\usepackage{listings}
\usepackage{url}

\def\BibTeX{{\rm B\kern-.05em{\sc i\kern-.025em b}\kern-.08em
    T\kern-.1667em\lower.7ex\hbox{E}\kern-.125emX}}

\begin{document}

\title{From Ad-Hoc Scripts to Orchestrated Pipelines: Architecting a Resilient ELT Framework for Developer Productivity Metrics}

\author{\IEEEauthorblockN{Yuvraj Agrawal, Pallav Jain}
\IEEEauthorblockA{\textit{Adobe Inc.}}
}

\maketitle

\begin{abstract}
Developer Productivity Dashboards are essential for visualizing DevOps performance metrics such as Deployment Frequency and Change Failure Rate (DORA). However, the utility of these dashboards is frequently undermined by data reliability issues. In early iterations of our platform, ad-hoc ingestion scripts (Cron jobs) led to "silent failures," where data gaps went undetected for days, eroding organizational trust. This paper reports on our experience migrating from legacy scheduling to a robust Extract-Load-Transform (ELT) pipeline using Directed Acyclic Graph (DAG) orchestration and Medallion Architecture. We detail the operational benefits of decoupling data extraction from transformation, the necessity of immutable raw history for metric redefinition, and the implementation of state-based dependency management. Our experience suggests that treating the metrics pipeline as a production-grade distributed system is a prerequisite for sustainable engineering analytics.
\end{abstract}


\section{Introduction}
\label{sec:introduction}

The measurement of software delivery performance using the DORA (DevOps Research and Assessment) framework has become a standard practice for high-performing technology organizations [1]. Metrics such as \textit{Deployment Frequency} and \textit{Lead Time for Changes} provide objective signals on engineering velocity and stability. However, while the definitions of these metrics are well-documented, the architectural patterns required to compute them reliably at scale are less explored.

In our organization's initial implementation, metrics were aggregated using isolated Python scripts scheduled via standard Cron utilities. This approach appeared sufficient for a single team but collapsed as we scaled to ingest data from Jira, GitHub, and Jenkins across multiple repositories and business units.

We encountered a critical data quality phenomenon which we formally define as the \textbf{Phantom Zero}:

\begin{quote}
\textit{\textbf{Phantom Zero:}} An failure mode in data pipelines where an upstream extraction error (e.g., a silent API pagination failure or timeout) is interpreted by downstream transformation logic as a valid empty set. This results in the calculation and reporting of "zero activity" (e.g., 0 Deployments) rather than a system exception, making the failure indistinguishable from a day of low engineering activity.
\end{quote}

In our legacy system, Phantom Zeros often persisted for days before being detected manually, significantly damaging stakeholder trust in the dashboard. A dashboard that reports "0 incidents" is useless if the user cannot distinguish between a stable system and a broken sensor.

This paper presents an architectural experience report on solving this "Last Mile" problem of data trust. We describe our migration to an orchestrated ELT pipeline that eliminates Phantom Zeros through:
\begin{enumerate}
    \item \textbf{State-Based Execution:} Transformations run only when dependencies are strictly satisfied, utilizing "Hard Gates" to prevent data pollution.
    \item \textbf{Time-Travel Capability:} The ability to recompute historical metrics when business definitions change without re-fetching raw data from rate-limited APIs.
    \item \textbf{Auditability:} A clear lineage from raw JSON payloads to final aggregated metrics.
\end{enumerate}

The scope of this work is limited to batch-oriented organizational health metrics; real-time operational telemetry (e.g., server CPU usage) is outside the scope of this architecture.

\section{The Architectural Paradigm}
\label{sec:architecture}

Our system ingests data from three primary disparate sources: structured issue tracking data (Jira), semi-structured version control metadata (GitHub), and unstructured build logs (Jenkins). To manage this complexity, we adopted an ELT (Extract-Load-Transform) paradigm supported by a Medallion Architecture.

\subsection{The Case for ELT over ETL}
In our early ETL (Extract-Transform-Load) attempts, we transformed raw API responses into metric values immediately upon ingestion. This proved fragile due to the evolving nature of metric definitions. For instance, when the organization redefined "Change Failure" to exclude "P4 Incidents," the ETL model required re-fetching nine months of historical data from the Jira API. This operation triggered strict API rate limits and resulted in partial data gaps.

By shifting to \textbf{ELT}, we prioritize loading raw JSON payloads into storage immediately. Transformations occur downstream within the database engine. This decoupling allows us to iterate on metric definitions (the "Transform" step) using stored history, independent of the source system's availability or rate limits.

\subsection{Medallion Architecture Implementation}
We organize data into three distinct layers of refinement to ensure data quality and traceability.

\subsubsection{Bronze Layer (Raw Audit)}
The Bronze layer acts as an immutable, append-only store for raw data.
\begin{itemize}
    \item \textbf{Design:} We store the exact JSON payload returned by the source API, enriched only with ingestion metadata (`fetched\_at`, `source\_system`, `execution\_id`). No validation is performed at this stage to maximize ingestion throughput.
    \item \textbf{Retention Strategy:} We initially retained Bronze data indefinitely. However, we observed that debugging utility diminishes rapidly over time. We now enforce a 30-day retention policy, which covers approximately 99\% of debugging scenarios while controlling storage costs.
    \item \textbf{Operational Value:} The Bronze layer serves as the reliable fallback mechanism. In three separate instances of downstream logic errors, we were able to replay the Bronze data to correct the Silver layer without making a single external API call.
\end{itemize}

\subsubsection{Silver Layer (Standardization)}
The Silver layer is responsible for normalization and cleaning.
\begin{itemize}
    \item \textbf{Schema Enforcement:} Timestamps from diverse sources (e.g., Jenkins' Unix epoch vs. GitHub's ISO 8601) are converted to a uniform UTC `datetime` object.
    \item \textbf{Identity Mapping:} A significant challenge in DORA metrics is identity resolution. A user may exist as `jdoe` in GitHub but `jane.doe@company.com` in Jira. The Silver layer applies a mapping logic to resolve these disparate identities into a single `contributor\_id`.
    \item \textbf{Filtering:} Invalid records, such as aborted Jenkins builds or "bot" commits, are filtered out at this stage.
\end{itemize}

\subsubsection{Gold Layer (Business Aggregates)}
The Gold layer is optimized for read performance and consumption by the dashboard visualization tool (e.g., Metabase or Tableau).
\begin{itemize}
    \item \textbf{Pre-Aggregation:} Instead of storing individual events, this layer stores aggregated metrics (e.g., "Daily Deployment Count," "Median Lead Time").
    \item \textbf{Query Latency:} Dashboard queries against the Gold layer scan hundreds of documents, whereas queries against the Silver layer would require scanning tens of thousands of records. This decoupling ensures dashboard responsiveness is not linear to the volume of raw data.
\end{itemize}

\begin{figure}[t]
    \centering
    \includegraphics[width=0.5\columnwidth]{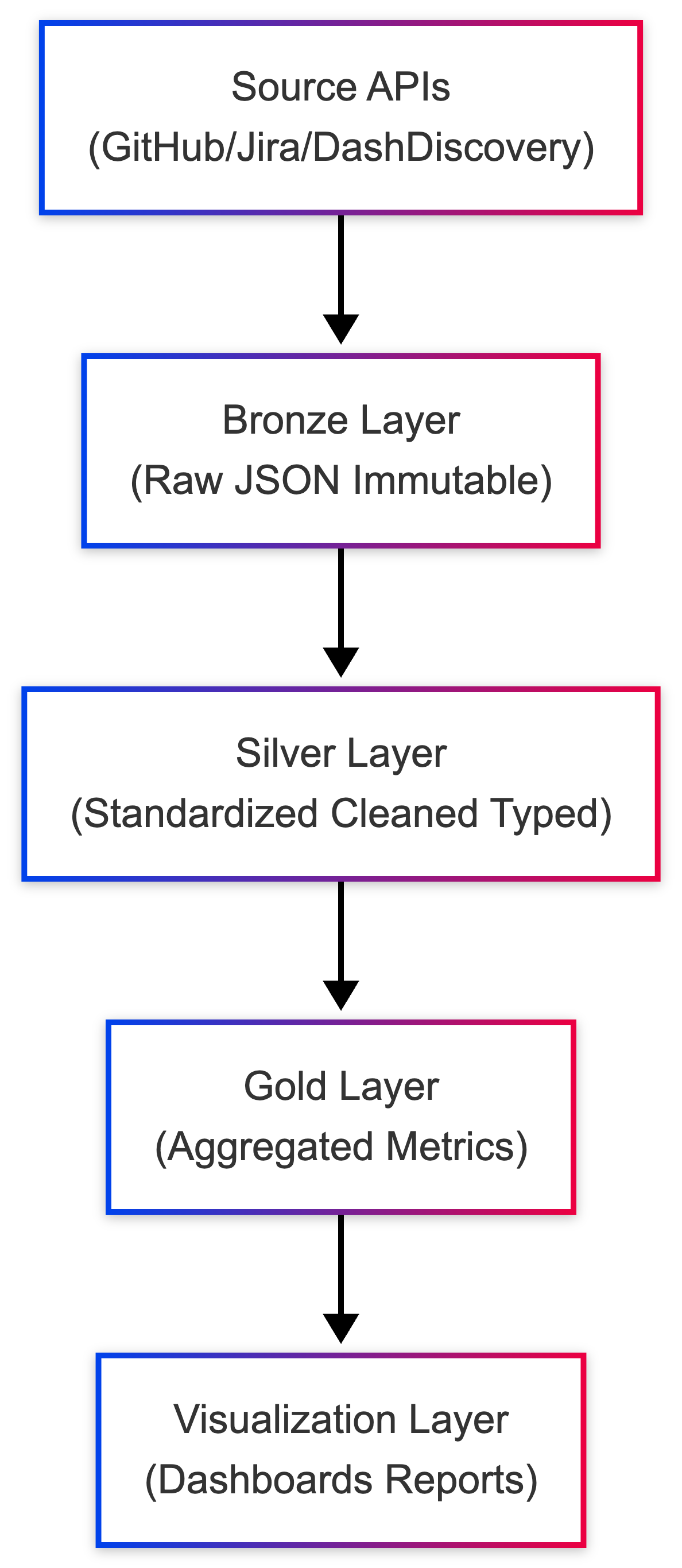}
    \caption{The Medallion Architecture applied to Developer Productivity Metrics.}
    \label{fig:medallion}
\end{figure}

\subsection{Storage Strategy: Time-Series Optimization}
Given the temporal nature of DORA metrics, we utilize MongoDB Time-Series collections (version 5.0+).
We observed that the choice of the `metaField` (partitioning key) is the single most critical design decision.
\begin{itemize}
    \item \textbf{Initial Failure:} We initially used `repository\_name` as the metadata field. However, in a microservices environment with hundreds of small repositories, this led to high cardinality and bucket fragmentation, degrading compression ratios.
    \item \textbf{Optimized Approach:} We refactored to use `team\_id` or `product\_area`. This aligned storage locality with query patterns, as dashboards are typically filtered by team, resulting in a 40\% improvement in query performance.
\end{itemize}

\section{Orchestration Strategy}
\label{sec:orchestration}

The structural layers described above are managed by an Orchestration Engine (Apache Airflow), modeling the pipeline as a Directed Acyclic Graph (DAG). This shift from time-based to state-based execution was a major contributing factor to improved reliability.

\subsection{State-Based Dependency Management}
In our legacy Cron approach, scripts were loosely coupled by time (e.g., Script A runs at 1:00 AM, Script B at 1:30 AM). If Script A failed or ran long, Script B would execute on stale data.
In the DAG approach, the `Calculate\_Gold\_Metrics` task has a strict, blocking dependency on the success of `Normalize\_Silver\_Data`. If the GitHub API is unavailable and the Bronze ingestion fails, the pipeline halts immediately. This "Hard Gate" mechanism ensures that we never publish partial or corrupted data to the Gold layer, prioritizing data correctness over freshness.

\subsection{Idempotency and the "Backfill" Capability}
A requirement for a robust ELT pipeline is idempotency: the ability to run the pipeline multiple times for the same execution date without creating duplicate records. This is achieved via "Upsert" (Update or Insert) operations.
\begin{itemize}
    \item \textbf{Composite Keys:} We define unique keys at the database level for the Silver and Gold layers. For example, a Gold record is uniquely identified by `{date, team\_id, metric\_type}`.
    \item \textbf{Atomic Writes:} Writes to the Gold layer are atomic. If a re-run occurs, the existing record for that day is replaced.
\end{itemize}

This idempotency enables "Time Travel" or Backfilling. When we introduced a new metric ("Mean Time To Recovery"), we defined the logic in the DAG and triggered a backfill for the past 365 days. The scheduler spawned worker processes to recompute the Gold layer from the existing Silver data in parallel, completing a year's worth of analytics in under 20 minutes.

\subsection{Resiliency Patterns}
To handle the inherent instability of external APIs, we implemented several resiliency patterns within the DAGs:
\begin{enumerate}
    \item \textbf{Exponential Backoff:} Tasks utilize a retry policy with exponential backoff (e.g., 3 retries, doubling delay from 5 to 20 minutes). We capped the maximum delay to 45 minutes to prevent "zombie tasks" from blocking the queue indefinitely.
    \item \textbf{Dead Letter Queues (DLQ):} If a task fails after all retries, the payload metadata is sent to a DLQ for manual inspection, preventing the failure from crashing the entire orchestration server.
    \item \textbf{Sensor Tasks:} We utilize "Sensors" to verify data volume before processing. If the number of fetched Jira tickets is 90\% lower than the 30-day moving average, the pipeline pauses and alerts an engineer, preventing the "Phantom Zero" issue where a silent API failure mimics a day of zero activity.
\end{enumerate}

\section{Real-Time Alerting: The "Push" Model}
\label{sec:alerting}

While the DORA metrics themselves are calculated in batches (e.g., daily or every 12 hours), the *notification* of critical regressions must be timely. A dashboard is a passive tool; relying on engineers to proactively check it introduces significant latency between a metric regression and remediation. To bridge this gap, we implemented an active alerting layer.

\subsection{The Latency Gap in "Pull" Models}
Our legacy alerting system relied on a separate Cron job that polled the database every 60 minutes to check for threshold breaches (e.g., Change Failure Rate $> 15\%$).
\begin{equation}
    Latency_{alert} \approx T_{poll} + T_{processing}
\end{equation}
This "Pull" model was inefficient. It wasted computational resources querying the database when no data had changed, and it introduced an unavoidable latency window. In one incident, a surge in deployment failures at 2:05 PM was not detected until the 3:00 PM poll, delaying the incident response by 55 minutes.

\subsection{Event-Driven Architecture via Change Streams}
We shifted to an event-driven "Push" model using MongoDB Change Streams. This mechanism allows applications to subscribe to real-time data changes at the database level without the complexity of tailing the operation log (oplog).

\subsubsection{Architecture Components}
The alerting pipeline consists of three decoupled components:
\begin{enumerate}
    \item \textbf{The Producer (DAG):} The Airflow DAG writes a calculated metric to the Gold layer. This is an atomic write operation.
    \item \textbf{The Event Bus (Database):} The database emits a change event containing the document delta.
    \item \textbf{The Consumer (Monitor Service):} A lightweight, always-on microservice listens for `insert` or `replace` events on the Gold collection.
\end{enumerate}

\begin{figure}[t]
    \centering
    \includegraphics[width=0.5\columnwidth]{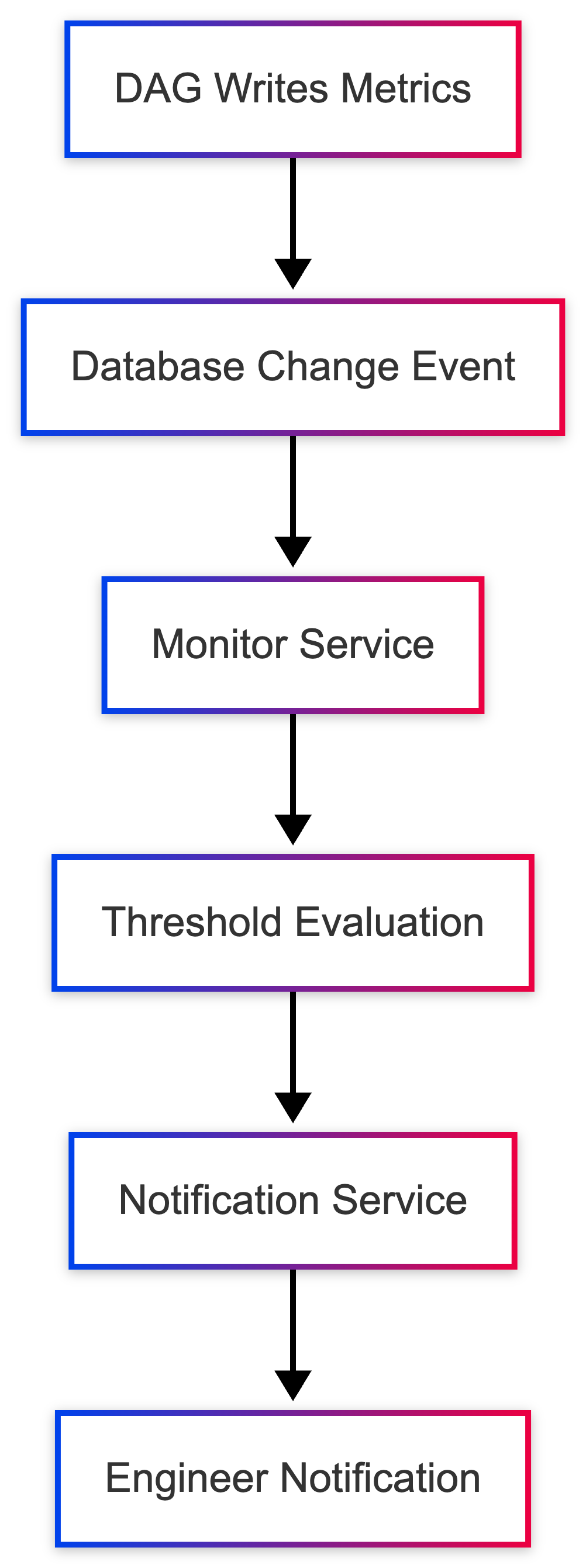}
    \caption{Real-Time Alerting Architecture via Change Streams.}
    \label{fig:medallion}
\end{figure}

\subsubsection{Resiliency via Resume Tokens}
A critical challenge in event-driven systems is reliability. If the Monitor Service crashes or is redeployed, events occurring during downtime could be lost.
We leveraged the `resume\_token` feature of Change Streams. The Monitor Service persists the token of the last successfully processed event to persistent storage (Redis). Upon restart, it presents this token to the database, which replays all events that occurred during the downtime.
\begin{itemize}
    \item \textbf{Observation:} This distinct delivery semantic ensures that no alerts are missed, even during system maintenance windows.
\end{itemize}

\section{Operational Impact and Observed Outcomes}
\label{sec:impact}

The migration from ad-hoc scripting to orchestrated ELT yielded measurable improvements in the operability and reliability of our metrics platform. Table \ref{tab:comparison} summarizes our observations comparing the legacy system to the new architecture over a 6-month operation period.

\begin{table}[htbp]
\caption{Operational Comparison: Legacy vs. Orchestrated}
\begin{center}
\begin{tabular}{@{}p{2.5cm}p{2.5cm}p{2.5cm}@{}}
\toprule
\textbf{Metric} & \textbf{Legacy (Cron)} & \textbf{Orchestrated (DAG)} \\
\midrule
\textbf{Failure Detection} & 2--5 Days (User reported) & $<$1 Hour (Automated) \\
\textbf{Data Lineage} & None (Black Box) & Full Task History \\
\textbf{Backfill Effort} & Manual Scripting (Days) & UI Trigger (Minutes) \\
\textbf{Recovery Time} & High (Manual Cleanup) & Low (Idempotent Rerun) \\
\bottomrule
\end{tabular}
\end{center}
\label{tab:comparison}
\end{table}

\subsection{Scenario Analysis: The Schema Change}
During a routine update, the Jira API schema changed, renaming a critical field used for "Lead Time" calculation.
\begin{itemize}
    \item \textbf{Legacy Outcome (Projected):} The script would have either crashed or, worse, inserted null values into the database. Recovery would have required writing a custom one-off script to delete the corrupted rows and re-fetch data.
    \item \textbf{Observed DAG Outcome:} The *Extract* task failed validation immediately. The pipeline halted, preventing any corruption of the Silver or Gold layers. The Data Engineering team was alerted via Slack. Once the parser was patched, we simply clicked "Clear State" in the Airflow UI. The pipeline resumed from the point of failure, automatically catching up on the missed data without duplicate ingestion.
\end{itemize}

\section{Lessons Learned}
\label{sec:lessons}

Throughout the design and implementation of this architecture, we identified several non-obvious principles that are critical for engineering analytics pipelines.

\subsection{Bronze is for Audit, Not Querying}
In our initial design, we attempted to build "live" dashboards directly on top of the Bronze (Raw) layer to reduce latency. This was a mistake. The schema inconsistency of raw API data led to complex, slow, and fragile queries that broke whenever a vendor changed their API response format.
\textbf{Lesson:} Strictly enforce the Medallion separation. The Gold layer creates a necessary contract between data engineering and data visualization.

\subsection{Alert on Absence of Data}
The most dangerous failure mode in analytics is not a crash, but a successful run that processes zero records (the "Phantom Zero"). Standard error monitoring does not catch this.
\textbf{Lesson:} We added "Data Volume Checks" as a final step in our DAGs. If the number of processed rows deviates by more than 2 standard deviations from the 30-day moving average, the pipeline alerts the team, even if the code executed without error.

\subsection{Token Rotation Management}
Hardcoded API tokens in environment variables proved to be a single point of failure and a security risk.
\textbf{Lesson:} We integrated our orchestration engine with a Secrets Manager. Tasks fetch fresh credentials dynamically at runtime. This significantly reduced authentication failures caused by expired tokens and improved our security posture.

\section{Limitations}
\label{sec:limitations}

While superior in reliability, the orchestrated ELT architecture introduces trade-offs that must be acknowledged.

\subsection{Infrastructure Complexity}
A Cron job requires a single server and a text file. An Airflow-based architecture requires a Scheduler, a Web Server, a Metadata Database (PostgreSQL), and a queuing mechanism (Redis). For small teams with low data volume and simple metrics, the operational overhead of maintaining this distributed system may outweigh the benefits of observability.

\subsection{Batch Latency}
Although the alerting layer is real-time relative to the write event, the metrics themselves are typically batch-processed (e.g., every 6 or 12 hours). The system reflects the state of the world as of the last DAG run. While sufficient for DORA metrics (which track long-term trends), this architecture is not suitable for real-time operational monitoring (e.g., server CPU spikes), which requires a streaming architecture (e.g., Apache Kafka or Flink).

\section{Future Work}
\label{sec:future}

While the current orchestrated ELT architecture has stabilized our reporting of DORA metrics, we have identified two primary areas for further evolution: the transition to streaming architectures for operational metrics and the adoption of machine learning for anomaly detection.

\subsection{Beyond Batch: Streaming Ingestion}
Our current Airflow implementation operates on a micro-batch schedule (e.g., every 4--6 hours). While sufficient for strategic metrics like \textit{Deployment Frequency}, this latency is suboptimal for operational metrics such as \textit{Time to Restore Service} during active incidents.
Future work involves implementing a hybrid architecture. We aim to introduce a streaming ingestion layer (using Apache Kafka or Confluent) alongside the batch layer. This would allow us to compute "Time to Recovery" in near-real-time using windowed aggregations (e.g., via Apache Flink), providing immediate feedback to incident commanders while retaining the batch layer for accurate historical trend analysis.

\subsection{AI-Driven Anomaly Detection}
Currently, our pipeline relies on static thresholds for data quality checks (e.g., the 90\% volume drop rule discussed in Section \ref{sec:orchestration}). While effective for catastrophic failures, static thresholds struggle with seasonality (e.g., holidays, weekends) and subtle degradations.
We are prototyping an unsupervised machine learning approach to learn the "normal" heartbeat of our engineering data. By replacing static rules with dynamic, seasonality-aware baselines, we aim to detect subtle data quality issues—such as a gradual drift in API latency or a slow decline in ingestion volume—before they trigger hard alerts, further reducing the cognitive load on the platform engineering team.

\section{Conclusion}
\label{sec:conclusion}

The transition from ad-hoc scripting to a disciplined ELT pipeline is not merely a technical upgrade; it is a prerequisite for building trust in Developer Productivity Dashboards. By adopting a Medallion Architecture, we ensure data auditability and consistency. By utilizing DAG-based orchestration, we eliminate silent failures and enable robust failure recovery. Finally, by layering event-driven Change Streams on top of the storage layer, we bridge the gap between historical analysis and real-time operational alerting.

Our experience suggests that the primary challenge in engineering analytics is not the definition of the metrics themselves, but the maintenance of confidence in the data supply chain that drives them. A dashboard that users do not trust is a dashboard that users do not use.

\appendices
\section{Airflow Dashboard Screenshots}

\begin{figure*}[p]
    \centering
    \includegraphics[height=0.25\textheight]{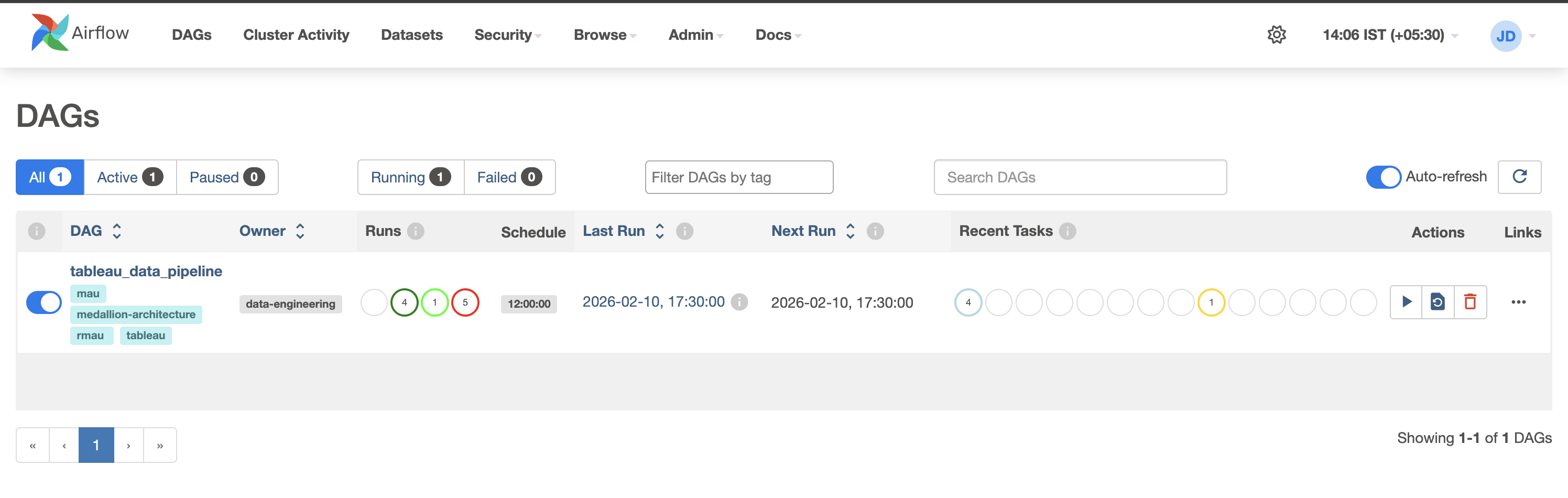}
    \caption{Airflow DAG Overview showing tasks and execution state.}
    \label{fig:airflow_dashboard_overview}
\end{figure*}

\begin{figure*}[p]
    \centering
    \includegraphics[height=0.35\textheight]{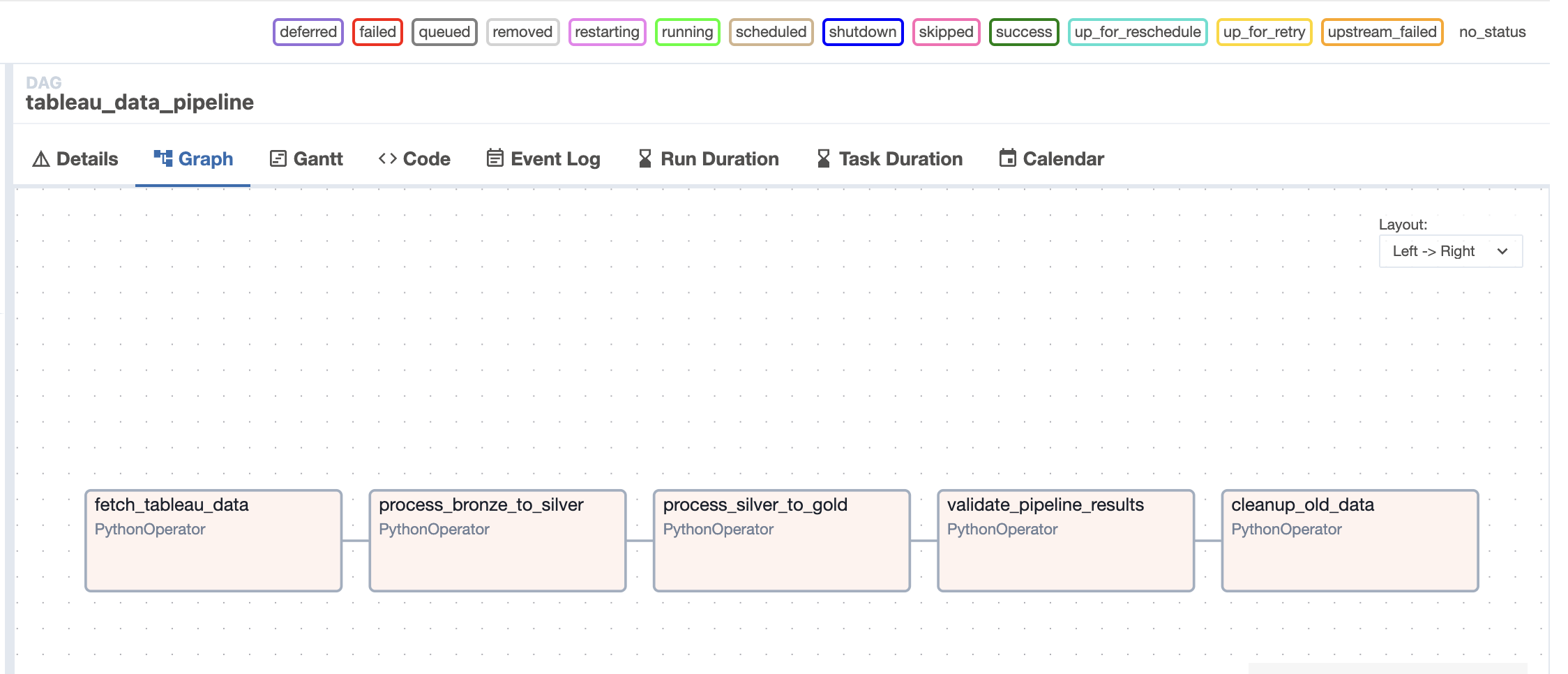}
    \caption{Airflow Graph View highlighting pipeline stages.}
    \label{fig:airflow_dashboard_taskview}
\end{figure*}

\end{document}